\documentclass[conference]{IEEEtran}
\IEEEoverridecommandlockouts
\usepackage{cite}
\usepackage{amsmath,amssymb,amsfonts,amsthm}
\usepackage{graphicx}
\usepackage{booktabs}
\usepackage{tikz}
\usetikzlibrary{automata, positioning, arrows.meta ,calc}
\usepackage[hidelinks]{hyperref}
\usepackage{textcomp}
\usepackage{xcolor}
\usepackage{algorithm}
\usepackage{algpseudocode}
\usepackage{pdfcomment}

\newtheorem{proposition}{Proposition}
\newcommand{\R}{\mathbb{R}}
\newcommand{\sat}{\operatorname{sat}}
\newcommand{\dd}{\mathrm{d}}
\newcommand{\pderiv}[2]{\frac{\partial #1}{\partial #2}}
\newcommand{\Vt}{V_{\mathrm{term}}}
\newcommand{\Vf}{V_{\mathrm{filt}}}
\newcommand{\dVf}{\dot{V}_{\mathrm{filt}}}
\newcommand{\Vi}{V_{\mathrm{init}}}
\newcommand{\Fon}{\mathrm{FracOn}}
\newcommand{\Vc}{V_{\mathrm{cease}}}
\newcommand{\Tc}{T_{\mathrm{cease}}}
\newcommand{\Td}{T_{\mathrm{delay}}}
\newcommand{\Vr}{V_{\mathrm{recon}}}
\newcommand{\Tr}{T_{\mathrm{recon}}}
\newcommand{\Tramp}{T_{\mathrm{ramp}}}
\newcommand{\Fc}{F_{\mathrm{cease}}}
\newcommand{\Fr}{F_{\mathrm{recon}}}
\newcommand{\MB}{\mathrm{MBase}}
\newcommand{\MWi}{\mathrm{MW}_{\mathrm{init}}}
\newcommand{\Mvi}{\mathrm{Mvar}_{\mathrm{init}}}
\newcommand{\Lfm}{L_{\mathrm{fm}}}
\newcommand{\Pctrl}{P_{\mathrm{ctrl}}}
\newcommand{\Qctrl}{Q_{\mathrm{ctrl}}}
\newcommand{\dP}{\mathrm{dP}}
\newcommand{\dQ}{\mathrm{dQ}}
\newcommand{\Ipv}{I_{pv}}
\newcommand{\Iqv}{I_{qv}}
\newcommand{\wf}{w_{\mathrm{filt}}}
\newcommand{\dwf}{\dot{w}_{\mathrm{filt}}}
\newcommand{\Kdroop}{K_{\mathrm{droop}}}
\newcommand{\Tsr}{T_{\mathrm{StartRamp}}}
\begin{document}

\title{A Hybrid Systems Approach to Parameterizing the PERC1 Model for Data Centers Using Synchrophasor Measurements}
\author{Bukunmi G. Odunlami,~\IEEEmembership{Student Member,~IEEE}, Marcos Netto,~\IEEEmembership{Senior Member,~IEEE}
\thanks{This work is supported by the National Science Foundation under Grant 2328241. The authors are with the Department of Electrical and Computer Engineering, New Jersey Institute of Technology, Newark, NJ 07102, USA.}
}

\maketitle

\begin{abstract}
The Power Electronic Reconnecting and Ceasing (PERC1) load model represents the timed ceasing and reconnection of large power-electronic loads, including data centers. Voltage thresholds and timers govern its response, with a ramp on reconnection, making it a hybrid dynamical system. This paper formulates PERC1 as a hybrid automaton and uses that structure to identify its parameters from synchrophasor records of grid disturbances. We propose a four-stage identification procedure: mode-sequence identification, estimation of logic parameters from event-time constraints, and joint Gauss--Newton refinement using hybrid trajectory sensitivities and saltation matrices. The resulting salted Fisher information is then used to analyze parameter identifiability. Two structural results are established. The detection and actuation delays are identifiable only through their sum unless a second dip interrupts the reconnection ramp, and a single disturbance leaves one unidentifiable direction in each threshold--delay pair, which two disturbances with different voltage slopes resolve. A case study using synthetic 60 frames/s synchrophasor records demonstrates recovery of the ceasing and reconnection dynamics and identifies poorly observable parameters.
\end{abstract}

\section{Introduction}
Data centers used 4.4\% of United States electricity in 2023, and projections for 2028 range from 6.7\% to 12\%~\cite{lbnl2024}. On July 10, 2024, a permanent 230~kV fault and three reclosing attempts sent about 1,500~MW of data center load to on-site backup supply. Most of it did not return for hours~\cite{nerc2025incident}. On July 22, 2026, a single 230~kV line fault in the Dominion zone of PJM removed about 3,800~MW in two waves one minute apart~\cite{pjm2026event}. In both events, the customers' own protection and controls initiated the disconnection. NERC's incident review therefore recommends that interconnection requirements include dynamic response models of large loads~\cite{nerc2025incident}.

The Power Electronic Reconnecting and Ceasing (PERC1) model was introduced in late 2025 to address this modeling gap~\cite{powerworld_perc1,weber2025}, and NERC recommends its use for representing the ceasing and reconnection of any type of data center~\cite{nerc2026trd}. Derived from the EPRI aggregated electric-vehicle charger model~\cite{epri2022}, it includes voltage-dependent load characteristics, current limits, washout and lead--lag blocks, and logic governing load ceasing and reconnection. The model reduces the connected load fraction when voltage remains below a specified threshold for a prescribed duration and ramps up following voltage recovery. 

The PERC1 parameters are generally unknown to system planners. Although uninterruptible power supply (UPS) vendors provide parameter ranges ~\cite{nerc2026trd}, these may not represent the response of individual facilities. To the best of our knowledge, no method has been proposed for jointly estimating PERC1's continuous and logic parameters from synchrophasor records while also analyzing their identifiability. Phasor measurement units (PMUs) at the point of interconnection provide disturbance records that could support such estimation. The challenge is to estimate the model parameters from these records and determine which parameters can be uniquely identified.

Measurement-based approaches have been widely studied for load model parameter estimation~\cite{hiskens2001,choi2006}, including PMU-based estimation~\cite{rouhani2016}. Gradient-based methods rely on smooth dependence of model outputs on the parameters. In PERC1, however, changes in the logic parameters alter when the load ceases and reconnection begins. Consequently, a least-squares objective evaluated using fixed-step simulations can be piecewise constant with respect to these parameters, limiting the effectiveness of gradient-based optimization. For the WECC composite load model, which also includes load tripping and reconnection, derivative-free optimization has been used with replayed or closed-loop simulations~\cite{woo2026}. However, such approaches do not establish which parameters can be identified from the available measurements, as illustrated by the difficulties in estimating trip thresholds reported in~\cite{woo2026}.

Hybrid dynamical systems offer a natural framework for addressing this problem~\cite{lygeros2003}. Hiskens and Pai \cite{hiskens2000} showed that the trajectory sensitivities of such systems jump at every transition and used them to fit models to disturbance recordings~\cite{hiskens2001}. The saltation matrix describes the jump~\cite{kong2024}, and the salted Fisher information framework of \cite{Odunlami2026} uses hybrid trajectory sensitivities to analyze parameter identifiability and establishes a hybrid persistence-of-excitation condition for local identifiability.

This paper develops a hybrid dynamical systems approach for estimating PERC1 parameters from synchrophasor measurements. We formulate PERC1 as a hybrid automaton and propose a four-stage identification procedure based on event-time constraints and hybrid trajectory sensitivities. Using salted Fisher information \cite{Odunlami2026}, we establish conditions for parameter identifiability and show how multiple disturbances can resolve ambiguities between voltage thresholds and timing parameters. 

This paper proceeds as follows. Section~\ref{sec:model} writes PERC1 as a hybrid automaton, and Section~\ref{sec:ident} builds the identification on it. Section~\ref{sec:case} tests the procedure on synthetic PMU records of a data center.

\section{PERC1 as a Hybrid Automaton}\label{sec:model}
In what follows, parameter and variable notation follows the vendor documentation~\cite{powerworld_perc1}. Quantities are in per unit on the model base $\MB = \MWi/\Lfm$, where $\MWi$ and $\Mvi$ are the pre-disturbance active and reactive power and $\Lfm$ is a loading factor~\cite{powerworld_perc1}. The inputs are the terminal voltage magnitude $\Vt$ and the frequency deviation $\Delta\omega$. The block diagram in~\cite{powerworld_perc1} has eight states: the filtered voltage $\Vf$ and frequency $\wf$, the washout outputs $\dP$ and $\dQ$, the lead--lag outputs $\Pctrl$ and $\Qctrl$, and the currents $I_p$ and $I_q$. We represent the washout and lead--lag blocks with the states $w_p$, $w_q$, $\ell_p$, $\ell_q$. Each washout is written in the partial-fraction form $sK/(1+sT) = (K/T)[1-1/(1+sT)]$, so the measured voltage is never differentiated, and each lead--lag as a feedthrough plus a low-pass state, $(1+sT_a)/(1+sT_b) = T_a/T_b + (1-T_a/T_b)/(1+sT_b)$. The corresponding continuous dynamics are given by
\begin{align}
T_v \dVf &= \Vt - \Vf, & T_f \dwf &= \sigma(\Vt)\Delta\omega - \wf, \label{eq:filt}\\
T_{vp}\dot w_p &= \Vt - w_p, & T_{vq}\dot w_q &= \Vt - w_q, \label{eq:wash}\\
T_{bp}\dot\ell_p &= u_p - \ell_p, & T_{bq}\dot\ell_q &= u_q - \ell_q, \label{eq:ll}
\end{align}
with outputs
\begin{align}
\dP &= \tfrac{K_{vp}}{T_{vp}}(\Vt-w_p), \quad \Pctrl = \ell_p + \tfrac{T_{ap}}{T_{bp}}(u_p-\ell_p), \label{eq:po}\\
\dQ &= \tfrac{K_{vq}}{T_{vq}}(\Vt-w_q), \quad \Qctrl = \ell_q + \tfrac{T_{aq}}{T_{bq}}(u_q-\ell_q), \label{eq:qo}
\end{align}
and lead--lag inputs $u_q = \Mvi/\MB + \dQ$ and $u_p = \MWi/\MB + \Kdroop\,\mathrm{db}(\wf) + \dP$, where $\mathrm{db}$ is the deadband $[\mathrm{dbfl},\mathrm{dbfh}]$ and the gate $\sigma(\Vt)$ in \eqref{eq:filt} is 1 while $\Vt\ge0.7$~pu and 0 otherwise. A voltage step thus produces a request pulse $K_{vp}\Delta\Vt/T_{vp}$ in \eqref{eq:po}. The requests are converted to currents with the filtered voltage, limited, scaled by the connected fraction $\Fon$, and delayed:
\begin{align}
\Ipv &= \sat_{[I_{p\min},I_{p\max}]}\!\big(\Pctrl\,(\Vf/\Vi)^{n_P}/\Vf\big), \label{eq:ip}\\
\Iqv &= \sat_{[I_{q\min},I_{q\max}]}\!\big(\Qctrl\,(\Vf/\Vi)^{n_Q}/\Vf\big), \label{eq:iq}\\
T_t\dot I_p &= \Fon\, \Ipv - I_p, \qquad P = \Vt I_p, \label{eq:it}\\
T_t\dot I_q &= \Fon\, \Iqv - I_q, \qquad Q = \Vt I_q, \label{eq:iq2}
\end{align}
where $P$ and $Q$ are $P_{\mathrm{MW}}$ and $Q_{\mathrm{Mvar}}$ on the base $\MB$~\cite{powerworld_perc1}. 

\subsection{Ceasing and reconnection logic}
The PERC1 logic is represented as a finite-state machine with modes Monitor, Cease, and Ramp~\cite{powerworld_perc1}. In Monitor or Ramp, if $\Vf$ stays below $\Vc$ for $\Tc$ seconds, a delay timer starts, and a recovery of $\Vf$ does not cancel it; after $\Td$ seconds, $\Fon$ drops to $1-\Fc$ and the mode becomes Cease. In Cease mode, once $\Vf$ has stayed above $\Vr\ge\Vc$ for $\Tr$ seconds, the time $\Tsr$ is recorded and $\Fon$ ramps linearly over $\Tramp$ seconds to $1-\Fc+\Fr\Fc$. The mode then returns to Monitor.

Subsequent ceasing events drop $\Fon$ to the same $1-\Fc$, so the share $\Fc(1-\Fr)$ leaves at the first cease and stays off, while $\Fr\Fc$ cycles with every event. A dip during the ramp goes through the same detector: $\Fon$ keeps ramping during the $\Tc$ dwell and freezes while the delay timer runs, then drops to $1-\Fc$.

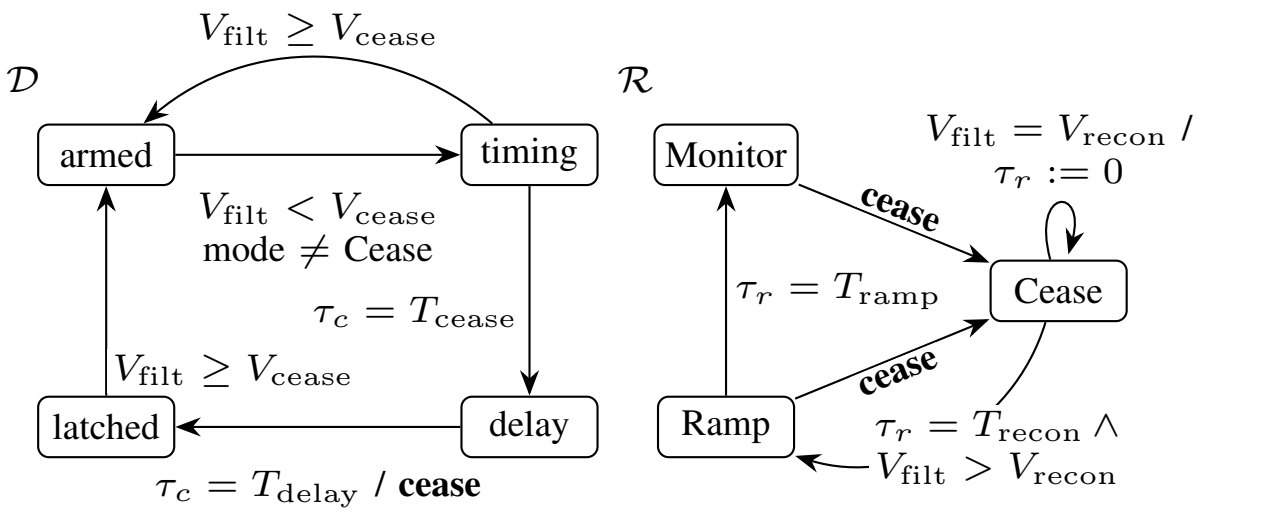
\begin{figure}[t]
\centering
\begin{tikzpicture}[>=Stealth, x=1cm, y=1cm,
  st/.style={draw, rounded corners=2pt, minimum height=4mm, minimum width=9mm, inner sep=2pt, font=\scriptsize},
  lab/.style={font=\scriptsize, inner sep=1pt, fill=white, align=center}]
\node[st] (d0) at (0,0) {armed};
\node[st] (d1) at (2.8,0) {timing};
\node[st] (d2) at (2.8,-1.8) {delay};
\node[st] (d3) at (0,-1.8) {latched};
\path[->]
 (d0) edge node[lab, below=6pt] {$\Vf<\Vc$\\ mode $\ne$ Cease} (d1)
 (d1) edge[bend right=40] node[lab, above=1pt] {$\Vf\ge\Vc$} (d0)
 (d1) edge node[lab, left=1pt, pos=0.6] {$\tau_c=\Tc$} (d2)
 (d2) edge node[lab, below=7pt] {$\tau_c=\Td$ / \textbf{cease}} (d3)
 (d3) edge node[lab, right, pos=0.12] {$\Vf\ge\Vc$} (d0);
\node[font=\scriptsize\bfseries] at (-0.55,0.5) {$\mathcal D$};
\node[st] (r0) at (4.1,0) {Monitor};
\node[st] (r3) at (4.1,-1.8) {Ramp};
\node[st] (r1) at (6.3,-0.9) {Cease};
\path[->]
 (r0) edge node[lab, sloped, above] {\textbf{cease}} (r1)
 (r3) edge node[lab, sloped, below] {\textbf{cease}} (r1)
 (r1) edge[bend left=45] node[lab, below, pos=0.3] {$\tau_r=\Tr\,\wedge$\\ $\Vf>\Vr$} (r3)
 (r1) edge[loop above] node[lab, above=2pt] {$\Vf=\Vr$ /\\ $\tau_r:=0$} (r1)
 (r3) edge node[lab, right=.5pt, pos=0.5] {$\tau_r=\Tramp$} (r0);
\node[font=\scriptsize\bfseries] at (3.5,0.5) {$\mathcal R$};
\end{tikzpicture}
\vspace{-.3cm}
\caption{Timed automata of the PERC1 logic. The arrows represent edges $E$, their transition conditions define guards $G$, and the state updates define resets $R$. $\mathcal D$ reads $\Vf$ from the plant \eqref{eq:filt}--\eqref{eq:iq2} and emits \emph{cease}, which resets $\Fon:=1-\Fc$ on both edges into Cease; the edge Ramp$\to$Monitor resets $\Fon:=1-\Fc+\Fr\Fc$. In Ramp, $\dd\Fon/\dd t=\Fr\Fc/\Tramp$ while $\mathcal D$ is not in delay, and $\Fon$ is constant otherwise. Clocks are reset to zero on entry to timing, delay, Cease, and Ramp, and $\tau_r$ is also reset by the self-loop of Cease.}
\label{fig:automata}
\end{figure}

We encode this as two timed automata~\cite{alur1994}
that share the event \emph{cease}
(Fig.~\ref{fig:automata}).
The cease detector $\mathcal D$ has clock $\tau_c$
and arms only while $\mathcal R$ is in Monitor or Ramp.
The mode automaton $\mathcal R$ has clock $\tau_r$
and operates over Monitor, Cease, and Ramp.
The transition from timing to armed resets $\tau_c$.
Entry to Cease and every crossing of $\Vr$ reset $\tau_r$,
so $\tau_r=\Tr$ with $\Vf>\Vr$ indicates
that the voltage has remained above the threshold
for the required duration.

\subsection{The hybrid automaton}
Let $x = (\Vf,\wf,w_p,w_q,\ell_p,\ell_q,I_p,I_q,\Fon,\tau_c,\tau_r)\in\R^{11}$ and $q = (d,r)\in Q = Q_{\mathcal D}\times Q_{\mathcal R}$. The PERC1 model is the hybrid automaton
$H = (Q,\R^{11},f,E,G,R)$ in the sense of~\cite{lygeros2003}. In mode $q$, the state flows according to $\dot x = f_q(x,u;\theta)$, given by \eqref{eq:filt}--\eqref{eq:iq2} with
\[
\frac{\dd\Fon}{\dd t}\in
\left\{0,\frac{\Fr\Fc}{\Tramp}\right\},
\qquad
\dot\tau_c,\dot\tau_r\in\{0,1\}.
\]
The edges $E$, guards $G$, and resets $R$ are shown in Fig.~\ref{fig:automata}. The parameter vector is divided into $\theta = (\theta_c,\theta_d)$. The continuous parameters $\theta_c$ include
gains, time constants, exponents, and current limits that govern the system dynamics. The logic parameters $\theta_d$ ($\Vc$, $\Tc$, $\Td$, $\Vr$, $\Tr$, $\Tramp$, $\Fc$, $\Fr$) appear only in the guards, resets, and ramp rate. Thus, the output $y = (P,Q)$ depends on $\theta_d$ through the transition times and the values of $\Fon$.

For a data center, $1-\Fc$ represents the load that rides through the disturbance (cooling, auxiliaries, and servers on static UPS units), $\Fc$ is the share transferred to backup supply, $\Fr\Fc$ is the share that automatically returns to grid supply, and $\Fc(1-\Fr)$ is the share that requires manual reconnection~\cite{nerc2025incident}.

\section{Identification from Synchrophasor Data}\label{sec:ident}

\subsection{Problem statement}
A PMU at the point of interconnection reports the positive-sequence voltage magnitude, frequency, active and reactive power at times $t_k = k/f_s$, with $f_s\in\{30,60\}$ frames/s~\cite{ieee_c37}: the inputs $\Vt$, $\Delta\omega$ and the outputs $y=(P,Q)$ of the model. We drive the model with the measured inputs, interpolated between frames, so the grid beyond the point of interconnection need not be modeled, as in the play-in approach of~\cite{hiskens2001,weber2025}. A phasor estimate is a windowed average of the underlying signal. We therefore apply the same window $\mathcal M_W$ to the model outputs before comparing them with the records $y_k$, and we discard samples within one window of a voltage step. The estimate minimizes, over all available records,
\begin{equation}
J(\theta) = \tfrac12\sum_{k} \big\| W_k^{1/2} r_k(\theta)\big\|^2 + \tfrac12\|\Sigma_0^{-1/2}(\theta-\theta_0)\|^2 , \label{eq:cost}
\end{equation}
where $r_k = \mathcal M_W y(t_k;\theta) - y_k$, $W_k$ weights by the inverse noise variances, and the second term is a weak prior centered at the initialization $\theta_0$ of Stages~1--2 below. It moves the estimate only along directions the data do not inform. 

\subsection{Hybrid trajectory sensitivities}\label{sec:sens}
Between transitions the sensitivity $S = \partial x/\partial\theta$ satisfies the variational equation $\dot S = (f_q)_x S + (f_q)_\theta$. At a transition at time $t_e$ defined by $g(x(t_e^-),\theta)=0$ with reset $x(t_e^+) = R(x(t_e^-),\theta)$, the event time depends on the parameters through~\cite[Prop.~3.4]{Odunlami2026}
\begin{equation}
\pderiv{t_e}{\theta} = -\frac{g_x S(t_e^-) + g_\theta}{g_x f^-}, \label{eq:te}
\end{equation}
and the sensitivity jumps as~\cite{hiskens2000}, \cite[Prop.~3.2]{Odunlami2026}
\begin{equation}
S(t_e^+) = \Xi\, S(t_e^-) + R_\theta - \big(R_x f^- - f^+\big)\frac{g_\theta}{g_x f^-}, \label{eq:salt}
\end{equation}
where $f^\mp$ are the vector fields before and after the transition and $\Xi = R_x - (R_x f^- - f^+)\,g_x/(g_x f^-)$ is the saltation matrix~\cite{kong2024}. The last term of \eqref{eq:salt} is the explicit dependence of the guard on the parameters, carried by $g_\theta$ in \eqref{eq:te}. For PERC1, it and $R_\theta$ are the only ways $\theta_d$ enters the output sensitivity. 

\begin{figure}[t]
\centering
\includegraphics[width=\columnwidth]{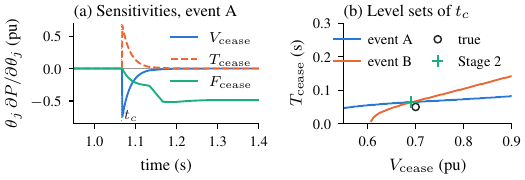}
\vspace{-.8cm}
\caption{(a) Normalized sensitivities $\theta_j\,\partial P/\partial\theta_j$ in event~A to $\Vc$, $\Tc$, and $\Fc$, from \eqref{eq:salt}. The timing parameters act through a pulse that decays with $T_t$; the fraction acts through a level change. (b) Level sets of the cease time in the $(\Vc,\Tc)$ plane for the two events.}
\label{fig:ident}
\end{figure}

For a clock guard $g = \tau-T$, $\partial t_e/\partial T = 1$. For a voltage guard $g=\pm(\Vf-V_\bullet)$, where $V_\bullet\in\{\Vc,\Vr\}$,  $\partial t_e/\partial V_\bullet = 1/\dVf(t_e)$. Thus, the slope of the filtered voltage at the threshold crossing determines how threshold changes affect the transition time. At a cease from Monitor, the only nonzero rows of $R_xf^- - f^+$ are those of $I_p$ and $I_q$, with values $\Fc\,\Ipv/T_t$ and $\Fc\,\Iqv/T_t$ respectively. Consequently, each timing parameter produces a pulse in $\partial P/\partial\theta$ of area $\Vt\Fc\Ipv\,\partial t_e/\partial\theta$ that decays with $T_t$ (Fig.~\ref{fig:ident}(a)). We integrate $S$ along with the state and apply \eqref{eq:salt} at each crossing, which we locate by interpolation within the step.

\subsection{Four-stage procedure}
\emph{Stage 1: Mode sequence.} Following~\cite{paoletti2007}, we first identify the mode sequence. For a single sag, the mode sequence is Monitor$\to$Cease$\to$Ramp$\to$Monitor. The connected fraction $\Fon$ follows the template $\Fon_T(t;t_c,t_r,t_e,a,b)$: 1 before $t_c$, $a$ until $t_r$, linear until $t_e$, and $b$ afterwards. The template is fitted to the ratio of the measured active power to the active power predicted by the model with $\Fon=1$ (Algorithm~\ref{alg:ident}, lines 1--3). The levels give $\Fc = 1-a$ and $\Fr = (b-a)/\Fc$, while the times give $\Tramp = t_e - t_r$, with $t_r$ corresponding to the vendor's $\Tsr$.

\begin{algorithm}[t]
\caption{Four-stage identification of PERC1}
\label{alg:ident}
\footnotesize
\begin{algorithmic}[1]
\Require Records $r=1,\dots,R$ of $\Vt$, $\Delta\omega$, $y_k=(P_k,Q_k)$; vendor $\theta_c^{\mathrm{nom}}$, initial $\theta_c^0$, prior $\Sigma_0$
\State \textit{Stage 1}, each record $r$: $\rho_k\gets P_k/\mathcal M_WP(t_k;\theta_c^{\mathrm{nom}},\Fon\equiv1)$
\State Fit $\Fon_T$ to $\{\rho_k\}$ on the frame grid by least squares, linear in $(a,b)$, to obtain $(t_c^{(r)},t_r^{(r)},t_e^{(r)},a^{(r)},b^{(r)})$
\State $\Fc\gets1-\bar a$, $\Fr\gets(\bar b-\bar a)/\Fc$, $\Tramp\gets\overline{t_e-t_r}$ (means over $r$)
\State \textit{Stage 2:} $t_\downarrow^{(r)}(V)$, $t_\uparrow^{(r)}(V)$ (after $t_c^{(r)}$) of $\Vt$ filtered with $T_v$
\State $\Vc\gets\arg\min_V\,[\max_r-\min_r]\big(t_c^{(r)}-t_\downarrow^{(r)}(V)\big)$ over $V$ with nonnegative delays (vendor default if $R=1$); $\Tc\gets$ mean of $t_c^{(r)}-t_\downarrow^{(r)}(\Vc)$; $\Td\gets0$
\State estimate $\Vr$, $\Tr$ from $t_r^{(r)}-t_\uparrow^{(r)}(V)$; $\theta\gets\theta_0:=(\theta_c^0,\theta_d)$
\Statex \textit{Stage 3} ($\lambda\gets10^{-2}$)
\Repeat
  \State for each $r$, integrate $\dot x=f_q$, $\dot S=(f_q)_xS+(f_q)_\theta$, applying \eqref{eq:te}--\eqref{eq:salt} at each guard crossing (located by interpolation)
  \State $r_k\gets\mathcal M_Wy(t_k;\theta)-y_k$, $J_k\gets\mathcal M_W\,\partial y(t_k)/\partial\theta$; $H\gets\sum_kJ_k^\top W_kJ_k+\Sigma_0^{-1}$, $g\gets\sum_kJ_k^\top W_kr_k+\Sigma_0^{-1}(\theta-\theta_0)$
  \State solve $(H+\lambda\,\mathrm{diag}\,H)\,\delta=-g$; if $J(\theta+\delta)<J(\theta)$ in \eqref{eq:cost}: $\theta\gets\theta+\delta$, $\lambda\gets\lambda/3$; else $\lambda\gets4\lambda$ and re-solve
\Until{the relative decrease of $J$ is below $10^{-6}$ or no step is accepted}
\State \textit{Stage 4:} $\mathcal F\gets\sum_kJ_k^\top W_kJ_k$ at $\hat\theta$; rank and null directions from the singular value decomposition of $W^{1/2}J$; if $\mathcal F\succ0$, $\sigma_j\gets\sqrt{[\mathcal F^{-1}]_{jj}}$
\end{algorithmic}
\end{algorithm}

\emph{Stage 2: guard parameters from event-time constraints.} Let $t_\downarrow(V)$ be the first time the filtered voltage of a record drops below $V$ and $t_\uparrow(V)$ the first time after the cease that it rises above $V$. A cease observed at $t_c$ and a ramp start observed at $t_r$ impose
{\setlength{\abovedisplayskip}{3pt}
\setlength{\belowdisplayskip}{2pt}
\begin{equation}
\begin{aligned}
t_c &= t_\downarrow(\Vc) + \Tc + \Td, \\
t_r &= t_\uparrow(\Vr) + \Tr,
\end{aligned}
\label{eq:constraints}
\end{equation}
}together with the dwell conditions that $\Vf$ stays below $\Vc$ on $[t_\downarrow, t_\downarrow+\Tc]$ and above $\Vr$ on $[t_\uparrow,t_\uparrow+\Tr]$. Each record contributes one equation per observed transition and an inequality for each dip below a candidate $\Vc$ without a cease. With two or more records, we select the threshold for which the delays implied by \eqref{eq:constraints} are consistent across the records.

\emph{Stage 3: joint refinement.} Starting from the estimates obtained in Stages~1--2, we use a Levenberg--Marquardt iteration to minimize \eqref{eq:cost} over $(\theta_c,\theta_d)$, with the Jacobian computed using \eqref{eq:salt} (Algorithm~\ref{alg:ident}, lines 7--11). 

\emph{Stage 4: identifiability.} The Gauss–Newton matrix $F=\sum_k J_k^\top W_k J_k$ is the salted Fisher information of\cite{Odunlami2026} for sampled outputs. Its rank and small singular values indicate which parameters can be identified from the records. When $F$ is positive definite,
the square root of the Cram\'er--Rao bound, $\sqrt{[F^{-1}]_{jj}}$,
provides a lower bound on the standard deviation of unbiased estimates of $\theta_j$.
\subsection{Two structural results}
\begin{proposition}\label{prop:delay}
For a record in which every cease is detected while $\mathcal R$ is in Monitor, the output of $H$ depends on $(\Tc,\Td)$ only through $\Tc+\Td$. A cease detected while $\mathcal R$ is in Ramp separates the two.
\end{proposition}
\begin{proof}
In Monitor, $\Fon\equiv1$ whatever the mode of $\mathcal D$, so the detector affects the plant only through the event \emph{cease}, emitted at $t_\downarrow(\Vc)+\Tc+\Td$. The delay mode has no exit guard on $\Vf$, so two parameter vectors with equal sum generate identical executions. In Ramp, $\dd\Fon/\dd t = \Fr\Fc/\Tramp$ while $\mathcal D$ is timing and $0$ while it is in delay, so the level at which $\Fon$ freezes, $1-\Fc+\Fr\Fc\,(t_\downarrow+\Tc-t_r)/\Tramp$, depends on $\Tc$ alone.
\end{proof}

For initialization, we set $T_{\mathrm{delay}}=0$ and assign the total delay to $T_{\mathrm{cease}}$; a second dip during ramp permits separate identification.

\begin{proposition}\label{prop:single}
For a record with a single cease at $t_c$ and a single ramp start at $t_r$, the Fisher information restricted to $(\Vc,\Tc)$ has rank one with null direction $(1,\,-1/\dVf(t_\downarrow))$, and restricted to $(\Vr,\Tr)$ has rank one with null direction $(1,\,-1/\dVf(t_\uparrow))$. Records of two disturbances whose filtered voltages cross the thresholds with different slopes make both blocks full rank.
\end{proposition}
\begin{proof}
By \eqref{eq:constraints}, the output depends on $(\Vc,\Tc)$ only through $t_c$ (the re-arm edges are unobservable without a second cease), so by \eqref{eq:te} the Jacobian columns satisfy $\partial y/\partial\Tc\,\delta\Tc + \partial y/\partial\Vc\,\delta\Vc = 0$ whenever $\delta\Tc = -\delta\Vc/\dVf(t_\downarrow)$; likewise for $(\Vr,\Tr)$ through $t_r$. Two records give two null directions that are independent when the slopes differ, so their intersection is trivial.
\end{proof}
In the language of \cite{Odunlami2026}, one record fails hybrid persistence of excitation along these directions, and a second record with a different crossing slope restores it.

\begin{figure*}[t]
\centering
\includegraphics[width=\textwidth]{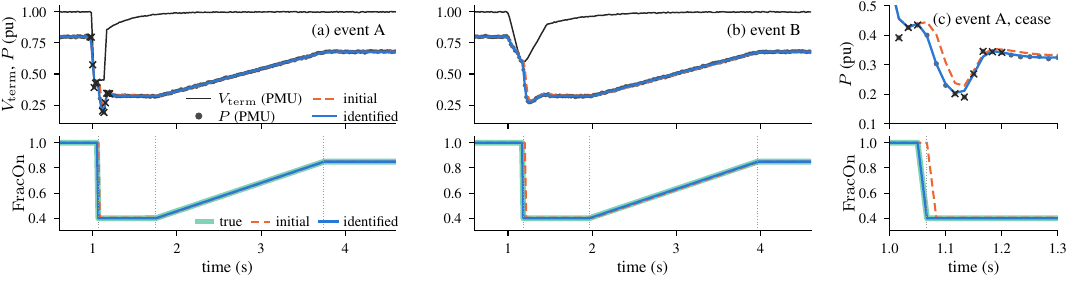}
\vspace{-.9cm}
\caption{Synthetic PMU records of (a) a nearby fault, (b) a remote slow sag, and (c) the cessation of event~A. Top: voltage and active power, measured and modeled at the Stage~1--2 initialization and after Stage~3; crosses mark masked samples within one phasor window of a voltage step. Bottom: $\Fon$, initial, true, and identified; dotted lines mark the true transitions.}
\label{fig:fit}
\end{figure*} 

\section{Case Study}\label{sec:case}
We consider a data center represented by PERC1 with the true parameters listed in Table~\ref{tab:results}. Following a sustained voltage sag below 0.70~pu for 50~ms, 60\% of the load transfers to backup. Of the disconnected load, 75\% reconnects after the filtered voltage remains above 0.88~pu for 0.5~s, with a 2~s ramp. The active power is nearly constant-power ($n_P=0.1$), with a current limit $I_{p\max}=1$. The frequency path is disabled ($\Kdroop=0$).

Two disturbances are considered. Event A represents a nearby fault, with voltage dropping to 0.45~pu for nine cycles, recovering to 0.85~pu at clearing, and returning to 1~pu with a 0.3~s time constant. Event B represents a remote disturbance, with voltage falling to 0.60~pu in 150~ms, holding for 60~ms, rising to 0.90~pu in 250~ms, and returning to 1~pu with a 0.5~s time constant. The model is simulated with a 1~ms step. Voltage and power are averaged over two cycles and sampled at 60~frames/s, with noise standard deviations of 0.002 and 0.003~pu, respectively.
\begin{table}[t]
\caption{Identification results (per unit, seconds)}
\label{tab:results}
\centering
\scriptsize
\vspace{-.3cm}
\setlength{\tabcolsep}{4pt}
\renewcommand{\arraystretch}{0.8}
\begin{tabular}{lccccc}
\toprule
Parameter & True & Stages 1--2 & Stage 3, A+B & CRB $\sigma$ & Stage 3, A only \\
\midrule
$K_{vp}$ & 0.020 & 0.010 & 0.017 & 0.015 & 0.027 \\
$T_{vp}$ & 0.100 & 0.100 & 0.115 & 0.44 & 0.107 \\
$K_{vq}$ & 0.010 & 0.005 & 0.011 & 0.0018 & 0.012 \\
$T_{vq}$ & 0.100 & 0.100 & 0.122 & 0.025 & 0.127 \\
$T_{ap}$ & 0.050 & 0.050 & 0.000 & 0.26 & 0.000 \\
$T_{bp}$ & 0.200 & 0.100 & 0.115 & 0.44 & 0.107 \\
$n_P$ & 0.100 & 0.000 & 0.096 & 0.029 & 0.146 \\
$n_Q$ & 1.000 & 1.000 & 0.991 & 0.035 & 1.067 \\
$I_{p\max}$ & 1.000 & 1.000 & 1.002 & 0.0030 & 1.002 \\
$T_v$ & 0.020 & 0.020 & 0.021 & 0.0076 & 0.023 \\
\midrule
$\Vc$ & 0.700 & 0.690 & 0.694 & 0.0083 & 0.520 \\
$\Tc$ & 0.050 & 0.064 & 0.047 & 0.0045 & 0.017 \\
$\Vr$ & 0.880 & 0.918 & 0.905 & 0.013 & 0.599 \\
$\Tr$ & 0.500 & 0.431 & 0.436 & 0.0088 & 0.578 \\
$\Tramp$ & 2.000 & 1.975 & 2.002 & 0.0069 & 2.007 \\
$\Fc$ & 0.600 & 0.595 & 0.601 & 0.0014 & 0.603 \\
$\Fr$ & 0.750 & 0.748 & 0.750 & 0.00069 & 0.751 \\
\bottomrule
\end{tabular}
\end{table}

Stage 1 identifies the cease times within approximately one PMU frame and estimates $F_{\mathrm{cease}}=0.595$, $F_{\mathrm{recon}}=0.748$, and $T_{\mathrm{ramp}}=1.975$~s. Stage 2 estimates the threshold--delay pairs from the observed transition times. As shown in Fig.~2(b), the intersection of the two level sets gives $(V_{\mathrm{cease}},T_{\mathrm{cease}})=(0.690,0.064)$, compared with the true values $(0.70,0.05)$. The reconnection parameters are estimated as $(V_{\mathrm{recon}},T_{\mathrm{recon}})=(0.918,0.431)$.

Stage~3 reduces the cost \eqref{eq:cost} from 4369 at initialization to 736 after three iterations and 725 after forty, compared with 751 at the true parameters and a noise floor of 710. The identified model (Fig.~\ref{fig:fit}) reproduces the cease times within 0.4~ms and the ramp start and end within 3~ms, below one PMU frame. Table~\ref{tab:results} reports the parameter estimates and Cram\'er--Rao standard deviations from the two-event Fisher information evaluated at the true parameters.

The logic parameters are recovered within two standard deviations, except $\Tr$, whose 64~ms error is seven times its bound. Over 20 noise realizations, the root-mean-square errors are 0.006~pu and 4~ms for $(\Vc,\Tc)$, 0.042~pu and 70~ms for $(\Vr,\Tr)$, 8~ms for $\Tramp$, and 0.003 and 0.001 for $\Fc$ and $\Fr$. The spread of $(\Vr,\Tr)$ exceeds the bound because input noise shifts the voltage threshold crossings, whereas the bound assumes exact voltage measurements. For $\Tramp$, $\Fc$, $\Fr$, and $I_{p\max}$, the spread is within 10\% of the bound, consistent with~\cite{Odunlami2026}. 

The current limit and voltage exponents are recovered, while $T_{vp}$, $T_{ap}$, and $T_{bp}$ remain poorly identifiable. During deep voltage sags, current limiting masks their effects, while phasor-window averaging smooths the short-duration responses needed to identify these time constants.

The single-event results in Table I confirm Proposition 2. For Event A alone, the Fisher information has rank 15 of 17, with null directions $(V_{\mathrm{cease}}, T_{\mathrm{cease}})\propto(0.997,0.079)$ and $(V_{\mathrm{recon}}, T_{\mathrm{recon}})\propto(0.522,-0.853)$, consistent with the crossing slopes of $-12.3$ and $0.55$ pu/s, respectively. Starting from the vendor default thresholds of 0.5 and 0.6 pu, the single-event estimates converge to $(V_{\mathrm{cease}}, T_{\mathrm{cease}})=(0.52,0.017)$ and $(V_{\mathrm{recon}}, T_{\mathrm{recon}})=(0.60,0.58)$. These estimates fit Event A as well as the true model but may predict incorrect transition times under other disturbances. Combining both events resolves the ambiguity, increasing the Jacobian's smallest singular value from $10^{-13}$ to $1.5$.

Proposition 1 is also verified by setting $T_{\mathrm{delay}}=50$ ms. For Event A, the Jacobian columns corresponding to $T_{\mathrm{cease}}$ and $T_{\mathrm{delay}}$ are identical, so the delays cannot be identified separately. Introducing a second voltage dip to 0.55 pu during the ramp makes the corresponding Jacobian block full rank, with singular values of 1662 and 27. The bound on $T_{\mathrm{delay}}$ is 31 ms, approximately two PMU frames. The 60 frames/s reporting rate limits the information available for estimating short detection delays, so the millisecond-scale detection delays reported by several vendors call for a PMU or fault recorder streaming at a higher rate.

\section{Conclusion}\label{sec:conc}
PERC1 is a hybrid system. Treating it as one turns its parameterization from curve fitting with a nonsmooth cost into a structured problem involving mode sequence identification, estimation of logic parameters from transition times, joint parameter estimation using hybrid trajectory sensitivities, and analysis of which parameters can be uniquely identified from the measurements using the Fisher information matrix. The hybrid formulation also makes explicit two facts that the block diagram hides. The detection and actuation delays cannot be identified separately unless a dip interrupts the reconnection ramp, while a single disturbance cannot uniquely identify the threshold-delay pairs. Two synchrophasor records of different voltage profiles recover the logic parameters of a data center load within the PMU frame resolution.  The washout and lead–lag time constants remain poorly identifiable because their effects are largely masked by current limiting during deep voltage sags.

\bibliographystyle{IEEEtran}
\bibliography{lib}

\end{document}